\documentclass[10pt, conference, letterpaper]{IEEEtran}

\ifCLASSINFOpdf
   \usepackage[pdftex]{graphicx}
\else
\fi
\newcommand{\subparagraph}{}

\usepackage{amsmath}

\ifCLASSOPTIONcompsoc
    \usepackage[caption=false, font=normalsize, labelfont=sf, textfont=sf]{subfig}
\else
\usepackage[caption=false, font=footnotesize]{subfig}
\fi
\begin{document}

%
\title{{\LARGE Optimizing IoT and Web Traffic Using Selective Edge Compression}}

\author{\IEEEauthorblockN{Themis Melissaris}
\IEEEauthorblockA{Princeton University\\
themis@cs.princeton.edu}
\and
\IEEEauthorblockN{Kelly Shaw}
\IEEEauthorblockA{Williams College\\
kshaw@cs.williams.edu}
\and
\IEEEauthorblockN{Margaret Martonosi}
\IEEEauthorblockA{Princeton University\\
mrm@princeton.edu}}

\maketitle


\begin{abstract}
Internet of Things (IoT) devices and applications are generating and communicating vast quantities of data, and the rate of data collection is increasing rapidly. These high communication volumes are challenging for energy-constrained,  data-capped,  wireless  mobile  devices and networked sensors.  Compression is commonly used to reduce web traffic,  to save  energy,  and to make  network  transfers  faster. If  not  used  judiciously, however, compression  can hurt performance. This work proposes and evaluates mechanisms that employ selective compression at the network's edge, based on data characteristics and network conditions. This approach (i) improves  the  performance  of  network  transfers in IoT  environments, while (ii) providing significant data savings. We demonstrate that our library  speeds up web  transfers by an average of 2.18x and 2.03x under fixed and dynamically changing network conditions respectively. Furthermore, it also provides consistent data savings, compacting data down to 19\% of the original data size.
\end{abstract}

\IEEEpeerreviewmaketitle

\setlength{\textfloatsep}{2pt}
\setlength{\belowcaptionskip}{1pt}
\setlength{\abovecaptionskip}{1pt}
\setlength{\floatsep}{0pt}
\section{Introduction}
Internet of Things (IoT) environments include elements for sensing, actuation, and communication, as well as data analysis and computation. IoT ecosystems promise to change the ways we interact with our homes and cities and to provide new solutions in industrial settings as well. As domains continue to benefit from nascent IoT technologies, they further contribute to the expanding availability and diversity of IoT devices. 

Increases in edge device counts, improved network infrastructure and the broad adoption of services and applications have led to an explosion in mobile and IoT data traffic, which is expected to increase nearly threefold from 2015 to 2020 \cite{cisco}. In fact, two-thirds of the total IP traffic by 2020 will be generated by wireless and mobile devices; this is due to an increase in the number of available connected mobile devices, adoption of wireless IoT devices, as well as growth in the devices' capabilities and data consumption. Energy is a primary constraint in designing applications and systems for edge devices and wireless communication accounts for a significant portion of the total energy budget, often dominating that of computation or other factors \cite{Carroll:2010:APC:1855840.1855861,Zhang:2004:HDE:1031495.1031522}.  With traffic and energy usage expected to surge, one technique proven to be efficient in managing the energy consumption and the traffic volume of wireless mobile devices is compression \cite{Sadler:2006:DCA:1182807.1182834}.

IoT's rise  has led to the broader emergence of an ecosystem of networked devices, supporting services, and new applications across many different domains \cite{iotapps}. Application areas such as smart surveillance, traffic services, and mobile sensing rely on data collected at the ``edge" (e.g. on smartphones or other mobile devices, rather than wired devices or cloud infrastructure) followed by communication of the data towards hub or cloud aggregators for analysis, often with tight latency requirements \cite{latency}.  
To support these data-intensive applications, our focus is on selectively using on-edge-device compression to reduce transferred bytecounts and improve communication efficiency.   

Data compression and decompression are widely available on commodity servers and can also be used at the edge to reduce the data exchanged in the network, sometimes reducing network latencies as well. Compression, however, needs to be used correctly to avoid overheads; if overused, it can {\em add} unnecessary latency and energy overhead to communications, instead of reducing them. Whether compression is beneficial or not is determined by several factors outlined below.

First, different mobile and IoT applications generate different types of content, which vary in size and compressibility.  The type of traffic generated on edge devices can change dynamically based on the users' interaction with the devices and the applications in use. Mobile web traffic is typically comprised of scripts, plaintext, multimedia and markup documents. IoT traffic includes sensor data that can vary significantly depending on the application and its usage. Variations in communicated data can mean significant variations in how compressible the data is.
For example, multimedia data items (e.g. audio, traffic and video) are usually already provided in a compressed format, preventing additional transfer-time compression from yielding large benefits.

Second, network behavior can significantly
alter the effect compression has on data exchanged over the network. In cases of low network throughput, compression can  reduce the duration of data transfers significantly. Alternatively, compression can stay in the application's critical path and introduce unnecessary overhead when the data compression rate is slower than the network data transfer rate. 

To selectively exploit the benefits of compression while intelligently avoiding its potential negative impact, this work proposes and evaluates the IoTZip approach.   IoTZip is a tool that allows mobile and IoT applications to handle compression intelligently. Based on characterizations of the data to be transferred and estimates of the network conditions, it  automatically reasons about compression trade-offs. It then predicts whether selective compression will pay off or not, and adaptively decides whether to use it to improve the performance of network transfers and reduce data usage. Our evaluation demonstrates that IoTZip achieves the stated goals in a very lightweight manner, which provides an opportunity for adoption of selective edge compression based approaches in resource constrained IoT environments.

As the edge increasingly includes data-intensive and latency-sensitive applications, the bandwidth and performance of wireless mobile devices become key design challenges. Intelligently compressing data going to and from IoT and wireless mobile edge devices can improve system functionality.

Our results show that IoTZip offers performance improvements of up to 3.78x (roughly 2x on average) and data size reductions of up to 81\%.  Interestingly, the IoT datasets we experiment on show more uniformity in size and data type than the mobile web datasets.  Nonetheless, selective compression is still useful even for predictable IoT datasets, because it allows systems to adapt to varying network conditions as well. 

The remainder of this paper is structured as follows.  Section II discusses related work, in order to further establish the motivation for IoTZip.  Section III describes the basic IoTZip functionality and Section IV gives the methodology and configuration information for its use in our experiments.  Section V presents our experimental results and Section VI offers conclusions.
\section{Related Work}
{\bf Characterizing Mobile Web Traffic \& Applications: }
One category of related prior work pertains to mobile web traffic characterization. The measurement study in \cite{Falaki:2010:FLT:1879141.1879176} discusses mobile traffic composition and investigates the performance and energy efficiency of TCP transfers. Butkiewicz et al. \cite{Butkiewicz:2011:UWC:2068816.2068846} studies parameters that affect web page load times across websites, whereas WProf \cite{Wang:2013:DPL:2482626.2482671} performs in-browser performance profiling. In \cite{Qian:2014:CRU:2594368.2594372}, the usage of bandwidth and energy in mobile web browsing is studied in detail using traffic collection and analysis tools, whereas \cite{Balasubramanian:2009:ECM:1644893.1644927} and \cite{mobconsumption} focus on analyzing the energy consumption of mobile devices' communication,  particularly mobile web browsing. In contrast to our work, these papers do not study compression, nor how the performance of a web transfer is affected by compressing data of varying data sizes and types.

{\bf Optimizing Mobile Web Traffic:}
Various techniques have been proposed to optimize mobile web transfers for performance and data usage. For example, Procrastinator \cite{Ravindranath:2014:VPP:2594368.2602432} decides when to prefetch objects in order to manage application data usage depending on a user's connectivity and data plan limitations. Other techniques like Polaris \cite{194916} and Shandian \cite{194914} use fine grained dependency tracking to identify and eliminate the intrinsic inefficiencies in the page load process. Klotski reprioritizes the delivery of web content in a dynamic fashion in order to improve the user experience \cite{klotski}.  While their focus is on improving performance and user experience in mobile web browsing, these approaches do not reduce data usage as we do. 

Compression-based approaches have also been proposed in related work to reduce data usage and improve performance. Locomotive \cite{locomotive} presents a methodology for determining at runtime whether compression is beneficial for data transmissions on mobile phones. Our work systematically studies the effect of compression on devices' web transfers at the Edge and builds a library that allows IoT applications to automatically adapt to dynamically changing network conditions and data heterogeneity. Additionally, compression proxies like Flywheel \cite{Agababov:2015:FGD:2789770.2789796}, Baidu TrafficGuard \cite{Li:2016:ECC:2930611.2930616} and Flexiweb \cite{Singh:2015:FNC:2789168.2790128} offer data savings by leveraging compression. These approaches, however, channel mobile content through a proxy server. Such rerouting raises privacy and security concerns if the proxy is untrusted and potentially latency concerns as well. Our work runs on mobile devices, performs compression at the Edge and therefore mitigates such concerns. 

Other works study how the use of different communication protocols affect the  performance of mobile web transfers. The study in \cite{spdy} compares HTTP 1.1 and SPDY (recently proposed HTTP alternative) performance in practice, showing no clear alternative advantage of the latter over cellular networks. In \cite{coap}, the authors present Lithe, a lightweight implementation of the CoAp protocol for the Internet of Things using compression to improve data usage and energy efficiency. This approach, despite its clear advantages, is restricted to low power wireless technologies (LoWPAN) and communication over CoAp. 
Recently, efforts in the industry has developed compression algorithms, like Brotli \cite{zstandard}, \cite{brotli}, \cite{lepton}, specifically designed for mobile traffic data savings and performance. 
Prior work has demonstrated that custom compression algorithms can achieve significant energy and performance gains \cite{Sadler:2006:DCA:1182807.1182834}. Using state of the art compression algorithms for IoT and mobile web traffic could complement network adaptive approaches such as IoTZip.

{\bf Correctness, Security and Privacy: } Related work also focuses on correctness, security and privacy aspects of IoT applications. OKAPI \cite{okapi} identifies correctness deficiencies and bugs in IoT applications and introduces tools that enforce correctness guarantees. Other works \cite{smartthings16} focus on security implications such as misuse of application privileges in IoT settings and develop privacy preserving solutions leveraging data protection and access control mechanisms \cite{flowfence16}.

{\bf Our Approach: } With IoTZip, application developers can optimize web traffic transfers from mobile and IoT devices through selective compression automatically. IoTZip dynamically decides whether to compress based on data characterizations and network conditions. IoTZip is device and application agnostic and therefore capable of enhancing all types of Internet of Things and mobile applications and benefit under heterogeneous traffic and under changing network conditions. Although IoTZip is provided as a library for application development, it can also be easily implemented as a browser plugin or extension. Section III  presents IoTZip's architecture, Section IV  describes our methodology and Section V  presents results on performance and data savings.

\section{IoTZip Library}
\begin{figure}[t]  
    \centering
    \includegraphics[width=0.5\textwidth]{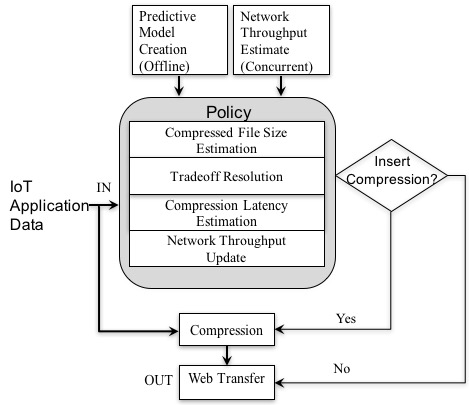}
    \vspace{-7pt}
    \label{fig:IoTZip}
    \caption{IoTZip Overview. Compression trade-offs are dynamically assessed during web transfers at the Edge. Whenever compression is estimated to be effective by the Policy module, data is compressed and transferred over the network. Network conditions are estimated using the Throughput Estimation module. The statistical models used for estimating compression time and size are generated offline in the Predictive Model Creation module. The tradeoff is resolved at runtime in the Tradeoff Resolution module.}
\end{figure}

\textbf{Overview}: IoTZip facilitates IoT and mobile traffic optimization by providing hooks allowing application programmers to use selective compression easily. Applications can invoke the library, abstracting away compression decisions. IoTZip focuses on uplink traffic where compression happens on the edge device. This is because uplink traffic is particularly latency- and energy-intensive for constrained IoT devices. IoTZip uses Android HTTP primitives, but can be extended easily to accommodate other protocols. In addition, a component that responds to IoTZip requests and handles data decompression runs in the cloud. The architecture of IoTZip is presented in Figure 1.   

\textbf{IoTZip Policy}: For all data transfers, IoTZip makes a two-step compression decision. First, a threshold determines if compression should be ruled out for some requests, based on size and compressibility. For small transfer sizes and for file types that are typically already compressed---such as multimedia---data compressibility can be low and the time spent compressing the data is likely to outweigh any benefits achieved. Selective compression avoids compressing in such cases.

IoTZip next determines if the estimated transfer latency is less with or without compression. As shown in Equation (I), IoTZip resolves the tradeoff for each request based on the compression latency $L_{Compression}$, the size of the request payload data before ($S_{Original}$) and after ($S_{Compressed}$) compression as well as the estimated network throughput, $N_{Throughput}$. $N_{Throughput}$ changes over time and is periodically estimated by the Network Connection Module. 
\begin{table}[t]
\centering 
\label{tab:equations}
\caption{Equation (I) resolves the performance tradeoffs of introducing compression. Equations (II) and (III) estimate data size after compression and compression latency, respectively.}
\begin{tabular}{ |l c| }
 \hline
(I) &$\frac{S_{Compressed}}{N_{Throughput}}+ L_{Compression} > \frac{S_{Original}}{N_{Throughput}}$\\
 \hline
(II) & $ S_{Compressed}( S_{Original}, T ) = \frac{S_{Original}}{compressibility(T)}$\\
\hline
 (III) & $ L_{Compression}( S_{Original}, T ) = \alpha(T)\cdot S_{Original} + \beta(T)$\\
\hline
i\end{tabular}
\end{table}

\textbf{Compression Size \& Time Estimation}: IoTZip estimates compression size and time using two  linear-regression-based statistical models whose input includes the data type $T$ and data size of the original data. To estimate the data size after compression, we use equation (II)'s model, where  $compressibility$ is the ratio between the data size of the original data versus the compressed. Equation (III) estimates the data compression latency, as a linear function of the data size and the coefficients $\alpha$, $\beta$; the coefficients are functions of the data type, and are acquired offline via training. To determine model parameters for equations (II), (III), IoTZip performs training during an initialization period that happens once at install-time. Since data patterns may vary over time, our model parameters can be updated with an online approach such as stochastic gradient descent using linear regression, but such adaptation is beyond the scope of this paper.

\textbf{Network Throughput Estimation}: IoTZip policy accounts for the dynamic behavior of network throughput $N_{Throughput}$ while selectively applying compression in network transfers. IoTZip samples the network periodically in order to provide accurate estimates in the face of fluctuations in network throughput, which achieves better accuracy. To acquire network throughput samples, IoTZip leverages an open source connection quality library and modifies the web server in the cloud, which are described in more detail in Section IV. 

\textbf{Tradeoff Resolution}: IoTZip puts all the estimates together in the Tradeoff Resolution
module to determine whether the web transfer savings of compressed data warrant incurring the additional compression latency. Once a decision has been reached, the data is compressed if necessary and a request is generated.
\section{Evaluation Methodology}\label{sec:meth}

\begin{figure}
\label{fig:size}
 \centering
  \includegraphics[width=1\linewidth]{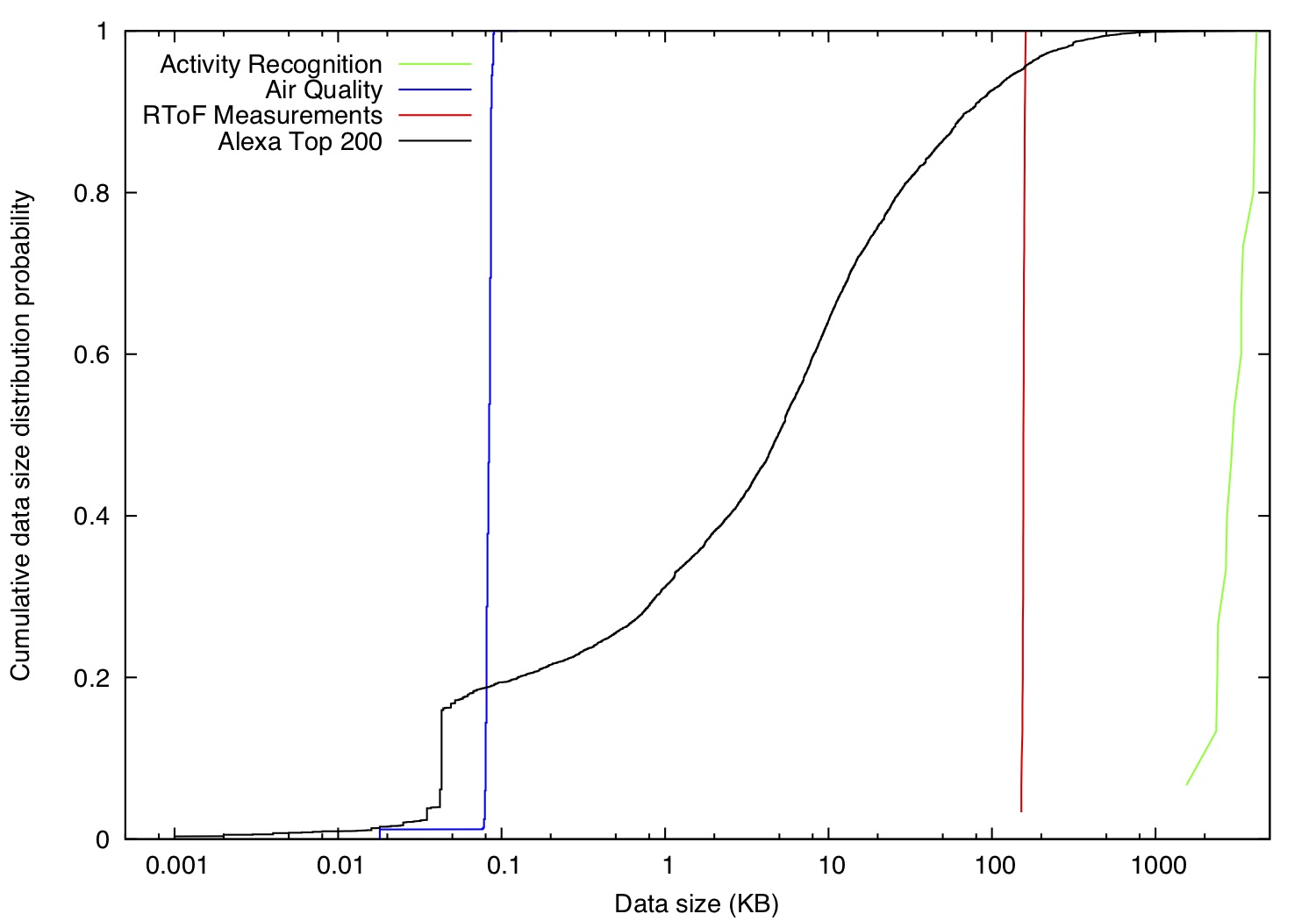}
  \vspace{-10pt}
   \caption{Cumulative Distribution Function (CDF) for the data size of all evaluated datasets. Alexa Top 200 is the most diverse in terms of data sizes, whereas the Air Quality and RToF Measurements are narrow and contain data small in size. Activity Recognition contains MB sized data.}
\end{figure}

\textbf{Edge implementation}: We evaluate our work using mobile web traffic, as well as IoT and sensor data found in related work. We replay web traffic on mobile phones using a test application that invokes IoTZip. IoTZip generates HTTP requests, which transfer web traffic data, as discussed later in this section. Once a compression decision has been made, the processed data will then be used to generate an HTTP request. IoTZip is intended to handle arbitrary data transfers. For the purpose of this work, we focus on HTTP, but other data transfer protocols could also benefit similarly. The client is implemented on Android and is run on a Samsung Galaxy S5 phone.

\textbf{Cloud setup}: To enable selective compression, our cloud infrastructure uses a web server capable of responding to HTTP requests. We vary network conditions in a controlled manner using Linux traffic shaping tools. The network throughput settings are 2 Mbps, 5 Mbps and 10 Mbps. 

\begin{table}[t]
\centering 
\label{tab:alexa}
\caption{Presentation of average data compressibility and data type distribution for four different Internet of Things and Mobile datasets.}
\setlength\tabcolsep{1.5pt}
\begin{tabular}{ || c | c | c | c || }
 \hline
 \textbf{Dataset}&\textbf{Data Type}& \textbf{Compressibility} & \textbf{Content Distribution}\\
 \hline
 Activity Recognition & Text &4.12 &100\%\\
 \hline
 Air Quality & Text &1.04 &100\%\\
 \hline
 RToF Measurements & Text &5.42 &100\%\\
 \hline
 Alexa Top 200 & Javascript &3.39 &38.36\%\\
 \hline
  & Images&1.02&32.07\% \\
 \hline
  & CSS &4.56 &10.25\%\\
\hline
 & HTML  &4.18 &7.32\%\\
\hline
 & Other &1.17 &8.74\%\\
 \hline
\end{tabular}

\end{table}

\textbf{Benchmarks}: We use three datasets to represent a range of IoT application domains. Activity Recognition \cite{dataset1} contains data taken from a wearable accelerometer and is collected from participants performing a range of activities. Air Quality \cite{dataset2} contains hourly responses of a gas multisensor device deployed on the field in a city in Italy. RToF Measurements  \cite{dataset3} includes Round-trip time-of-flight (RToF) and magnetometer measurements from 30 deployed stationary anchors in a supermarket indoors.  For all datasets, we consider scenarios where data across different networked sensors are transmitted from the devices and aggregated at the edge. 

In addition to the aforementioned datasets, we collected web traffic to replay on mobile platforms. Collection of the mobile web traffic was performed offline using Fiddler \cite{fiddler}, a web debugging proxy, which captures the raw payload of each request. To emulate real mobile traffic, we generate HTTP requests to transfer captured web traffic data as the request payload. We captured the traffic from mobile versions of the top 200 most popular websites according to the Alexa list \cite{Alexa}. For each web page load in the Alexa list, multiple (2 to 200) web page elements are fetched and loaded on the client. Although the traffic in the Alexa dataset is originating mostly from downlink traffic, we will consider that the data is living at the edge and will be using it for transfers between the edge and the cloud.  The dataset size is 350MB and consists of 25 different data formats, including scripts (e.g. HTML, Javascript), text formats (e.g. .txt files, JSON and XML formatted text) and multimedia (e.g. jpeg, png images, audio files). The Alexa Top 200 list contains a broad spectrum in terms of data size and data types. To the best of our knowledge, there are no alternative benchmarks available capable of capturing traffic representative of the wide range of mobile and Internet of Things devices.

To study IoTZip's behavior on different levels of compressibility, we created two test sets using traffic from 50 different websites. Test set A includes the 25 most compressible websites (average compressibility 3.07) and test B includes the 25 least compressible websites (average compressibility 1.23) of the Alexa Top 200 list. We eliminated web sites that were either very small in size (order of a few tens of Kilobytes) or contained a small number of files. The remaining 150 websites comprise our training set, which we use to train IoTZip's models. 

Figure 2 presents the Cumulative Distribution Functions (CDFs) for each of the evaluated datasets. Data found in the IoT datasets vary orders of magnitude in size across datasets, but insignificantly within the same dataset. Data size in the Activity Recognition dataset is in the order of MBs, in the RToF Measurements dataset data size is in the order of KBs , whereas Air Quality data are very small (order of bytes). Alexa Top 200 data sizes spread from bytes to MBs.

Table II characterizes the datasets per data type and focuses on data compressibility and content distribution by data size. Activity Recognition and RToF Measurements include highly compressible text data. The Air Quality dataset has very low compressibility despite having text data due to the small data size; the compression algorithm builds a dictionary that is comparable in size to the original data size. Alexa Top 200 is largely composed by scripts and text data which are highly compressible and by images that are previously encoded and therefore yield very low compressibility.

In the experiments performed, we compare IoTZip against other different compression policies using the aforementioned benchmarks. As the applications and the target hardware vary significantly, we are not using mobile web browsers for our evaluation; instead, we focus on the total time required for a benchmark to complete the transfer over the network. We also account for  compression and decompression latency at the endpoints. 

\textbf{Network Throughput Estimation}:
We use the open source Network Connection Class \cite{connectionclass}, an Android library that allows developers to determine current network throughput of an application. Network Connection Class achieves this by listening to the traffic flowing through the application and by measuring network throughput samples along the way. The library uses throughput samples to keep a moving average of the network throughput and provides the user with a notification when there is a significant change. 

\textbf{Performance Evaluation}: For performance evaluation, we compare IoTZip against (i) a policy that performs all data transfers uncompressed ({\em Uncompressed}), (ii) an approach that compresses all data before they get transferred ({\em Compressed}) and against an oracle ({\em Time Oracle}). The Time Oracle always makes a correct decision when reasoning about the compression decision as it is computed by choosing the minimum request latency between compressing data and leaving it uncompressed for each individual web data transfer.

For each dataset we evaluate, each individual file or data item is processed by IoTZip and eventually transferred from the client at the edge to the cloud using an HTTP request. 
The resulting time required to transfer the mobile web site is the aggregate time of individual data transfer times of its elements. 

\textbf{Testing under changing network conditions}: IoTZip is able to perform under varying network conditions. To evaluate our framework in a dynamically changing environment, we emulate a network whose bandwidth varies over time and experiment with IoTZip's capacity to adapt to changes. We generate traces that encapsulate changes in network conditions and that vary over the course of our experiment. Each of these network conditions corresponds to a fixed network throughput level that remains constant during an epoch, a predefined period of time. Throughout the experiments, network settings are controlled in the cloud setup by traffic shaping tools and  network throughput levels are sampled to vary in a uniformly random manner. Using this methodology, we can test across controlled but varying network conditions.

In order to vary network settings in discrete intervals, we create a 4-way partition of each dataset and consider the time required for each dataset partition's transfer to complete as an epoch. During the timeline of the experiment we monitor the percentage of data compressed for IoTZip and compare it against the Time Oracle. This comparison will provide us with insight into how IoTZip adapts to the network changes and whether it decides to compress data at a higher or lower percentage, depending on the data and the network conditions. In addition, we compare IoTZip's performance against the Uncompressed, Compressed and Time Oracle policies. 

\section{IoTZip Evaluation}

\begin{figure}
 \centering
  \includegraphics[width=1\linewidth]{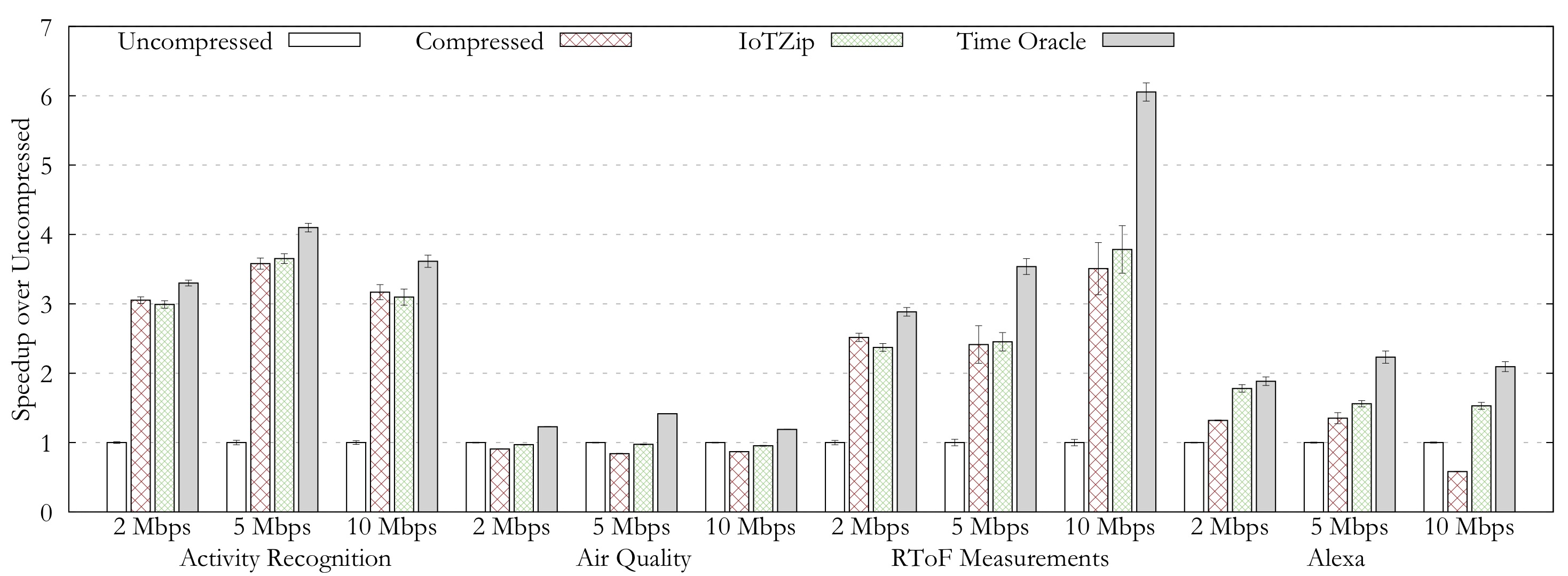}
  \label{fig:results}
  \vspace{-10pt}
   \caption{Speedup of compression policies across datasets and network settings. Results are presented relative to the Uncompressed policy along with their respective standard errors.}
\end{figure}

\begin{figure*}
  \centering
  \subfloat[High compressibility (A), 10 Mbps network bandwidth]{%
  \includegraphics[width=0.5\linewidth]{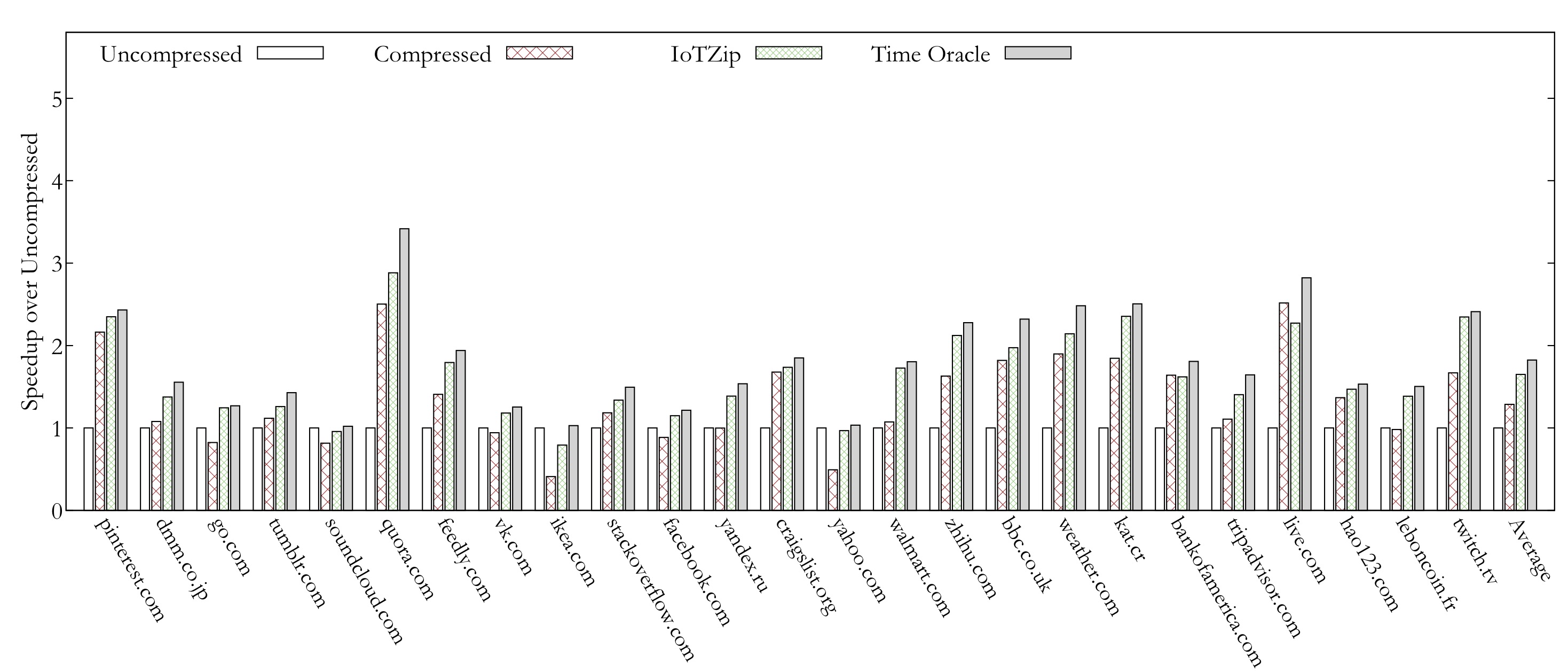}}
  \label{High compressibility (A), 10 Mbps network bandwidth}\hfill
 \subfloat[Low compressibility (B), 10 Mbps network bandwidth]{%
 \includegraphics[width=0.5\linewidth]{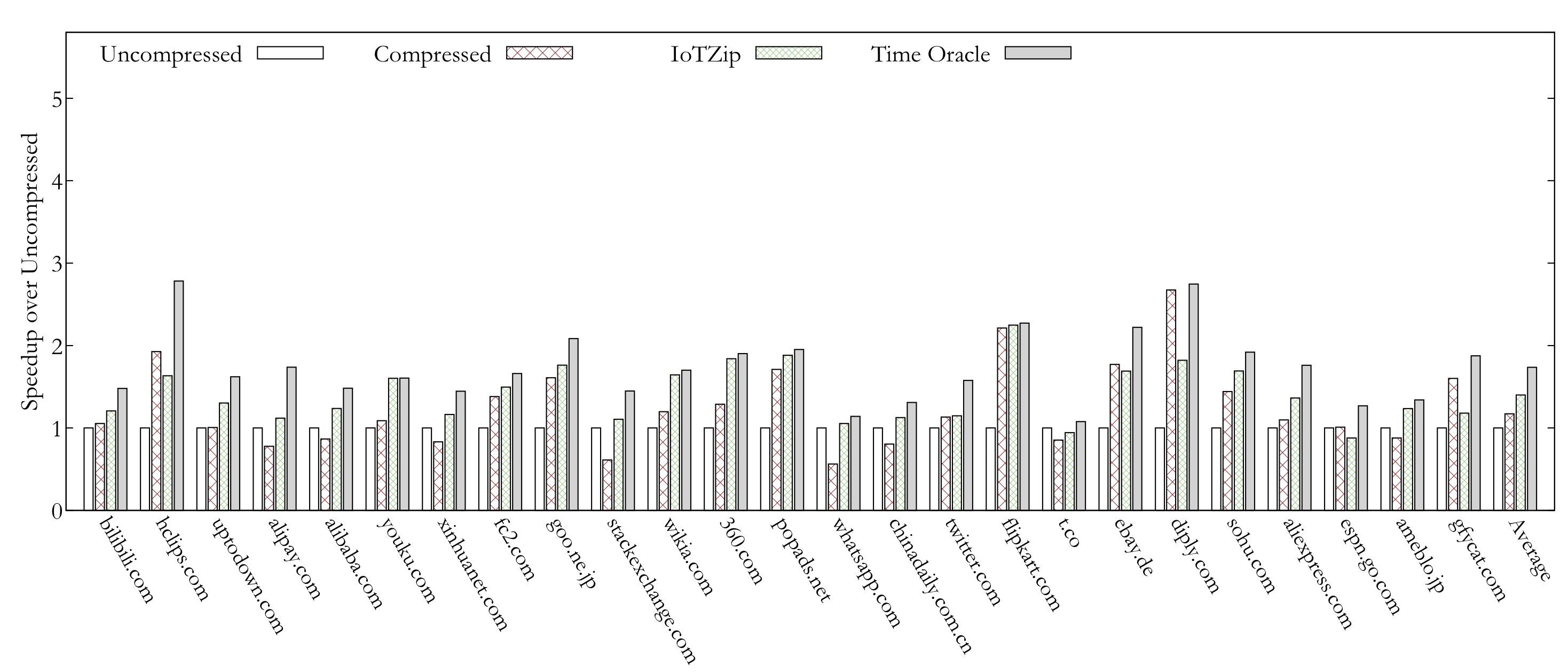}}
  \label{Low compressibility (B), 10 Mbps network bandwidth } \\
    \subfloat[High compressibility (A), 5 Mbps network bandwidth]{%
  \includegraphics[width=0.5\linewidth]{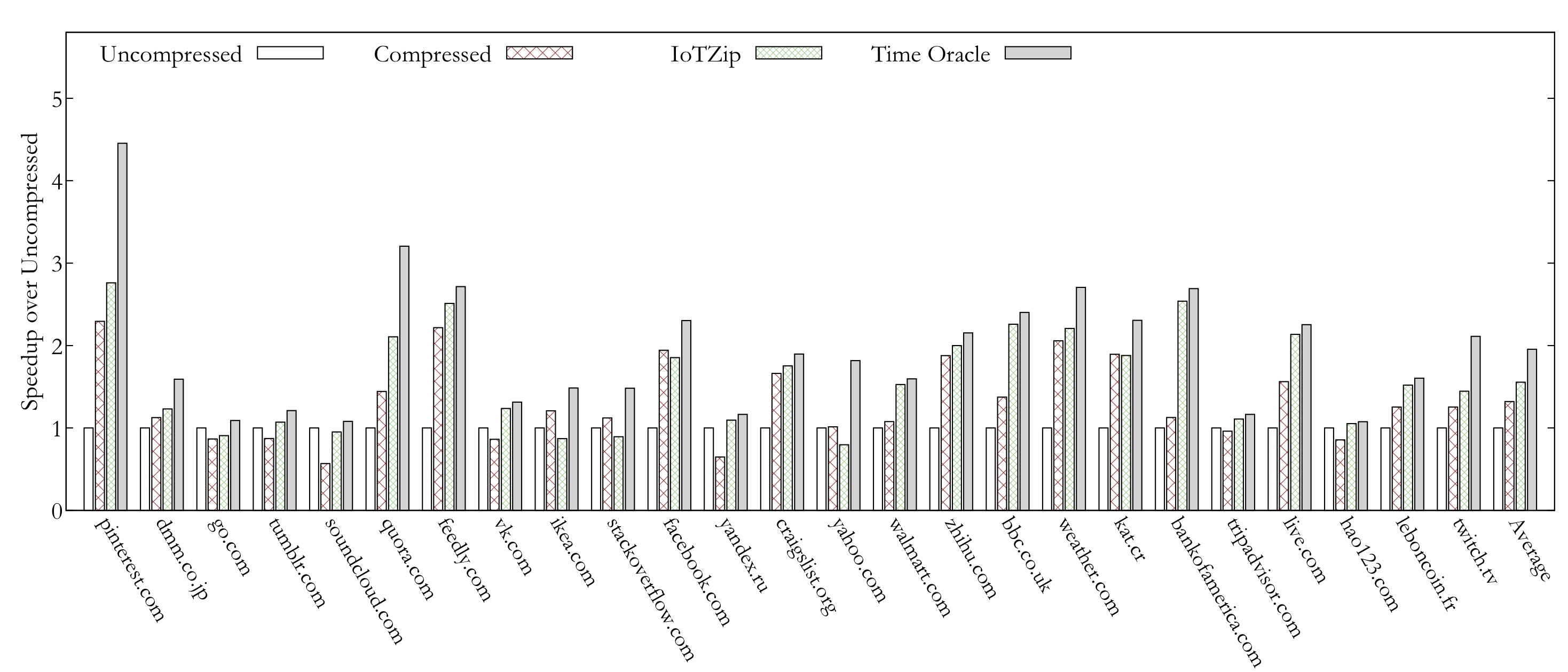}}
  \label{High compressibility (A), 5 Mbps network bandwidth}\hfill
 \subfloat[Low compressibility (B), 5 Mbps network bandwidth ]{%
 \includegraphics[width=0.5\linewidth]{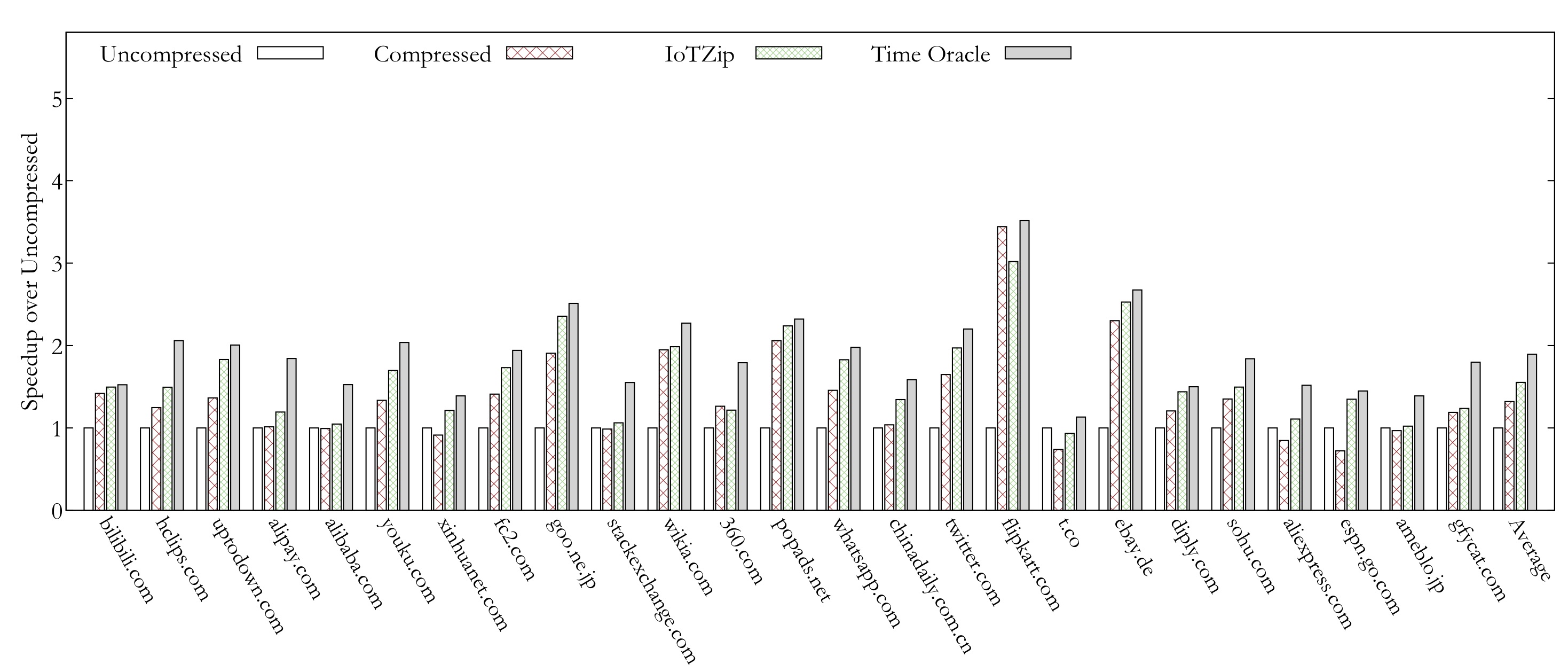}}
  \label{Low compressibility (B), 5 Mbps network bandwidth } \\
    \subfloat[High compressibility (A), 2 Mbps network bandwidth]{%
  \includegraphics[width=0.5\linewidth]{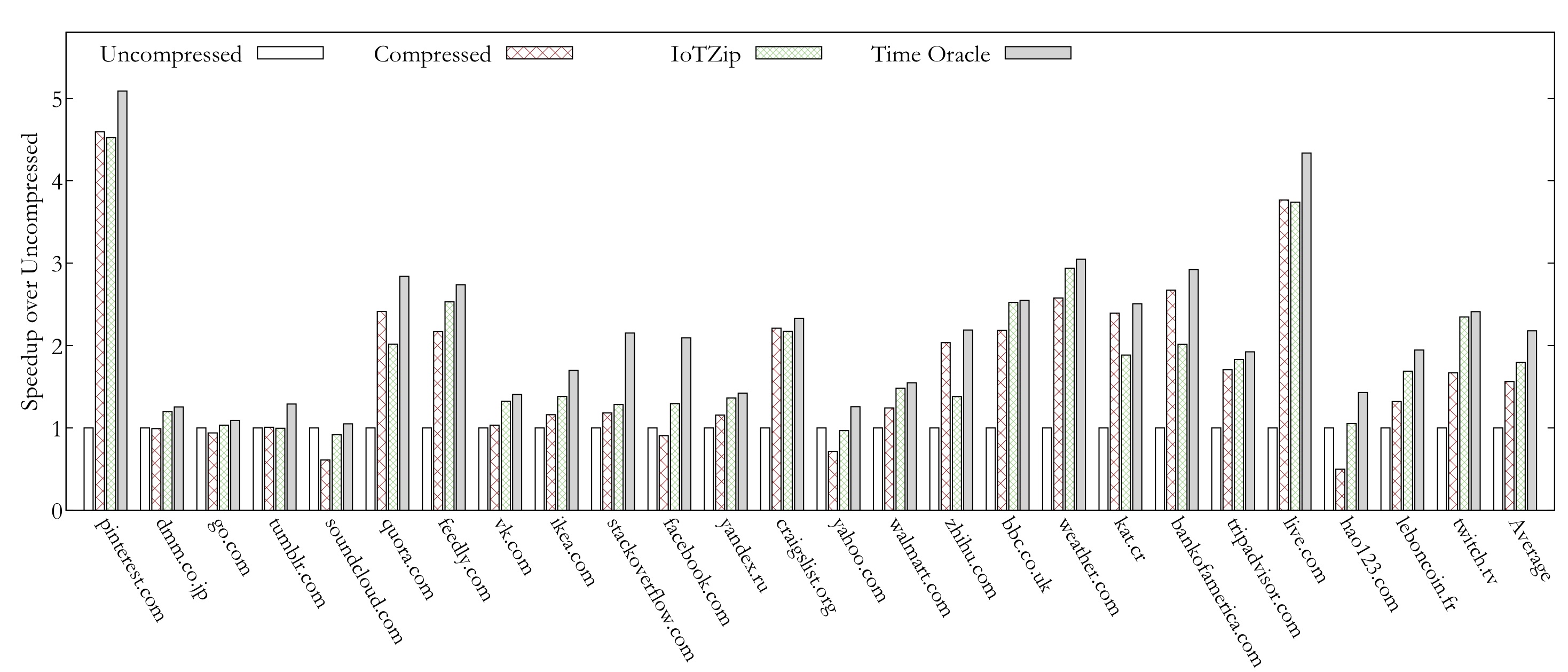}}
  \label{High compressibility (A), 2 Mbps network bandwidth}\hfill
 \subfloat[Low compressibility (B), 2 Mbps network bandwidth ]{%
 \includegraphics[width=0.5\linewidth]{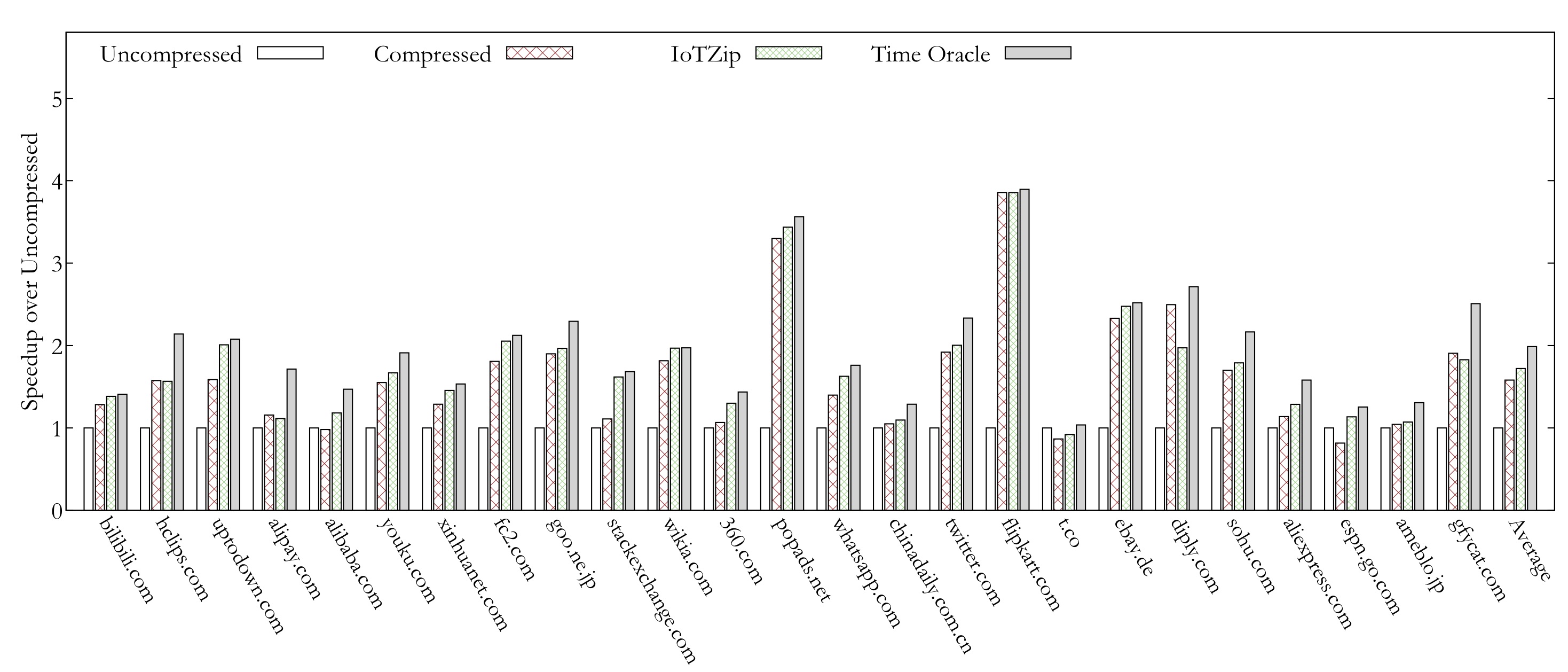}}
  \label{Low compressibility (B), 2 Mbps network bandwidth }\\
\caption{Performance evaluation across different network settings, using test sets of high compressibility (A) and low compressibility (B). ZipIoT is compared against an approach utilizing no compression (Uncompressed), an approach that compresses all data (Compressed) and a Time Oracle. For relative speedup comparison, higher is better, with an ideal Time Oracle representing a ``perfect'' performance. Comparison is available for network throughput set at 2,5,10 Mbps. Regardless of compressibility or network conditions, IoTZip demonstrates speedup against non-Oracle approaches.}
\label{fig:results}
\end{figure*}
\begin{figure}
 \centering
  \includegraphics[width=1\linewidth]{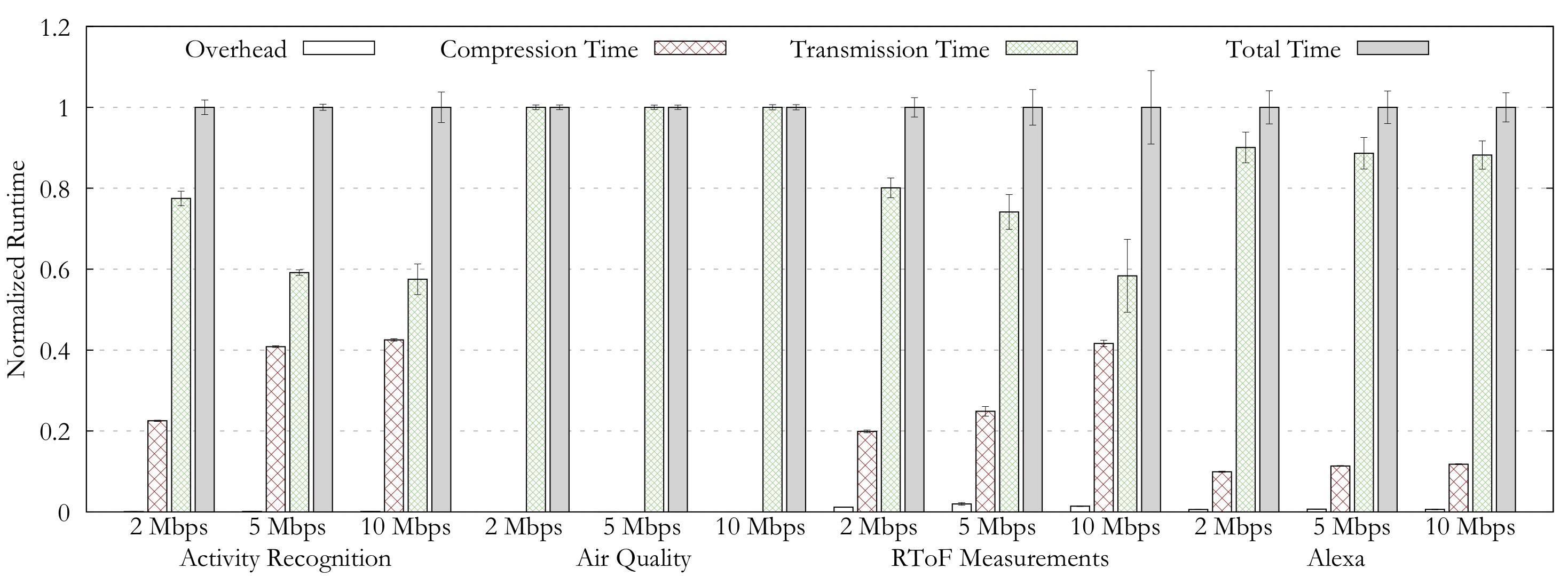}
  \label{fig:results}
  \vspace{-10pt}
   \caption{Breakdown of IoTZip's Total Runtime into Overhead, Compression time and Transmission Time. Results are normalized against the Total Runtime along with their respective standard errors.}
\end{figure}

\begin{figure}
 \centering
  \includegraphics[width=1\linewidth]{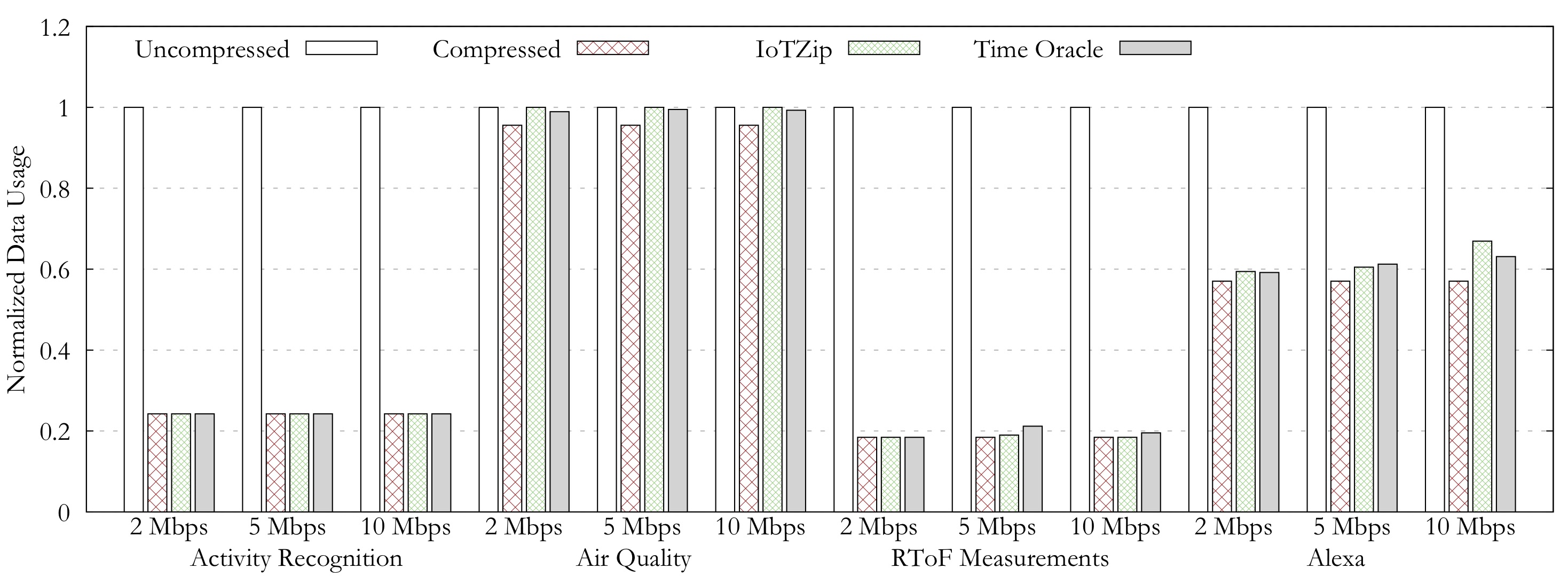}
  \label{fig:results}
  \vspace{-10pt}
   \caption{Presentation of Data Usage across datasets and network settings. Data Usage is normalized against the Uncompressed policy. Lower is better.}
\end{figure}

This section compare IoTZip's performance  against the (i) Uncompressed, (ii) Compressed and (iii) Time Oracle approaches previously described. In addition, we present statistics that showcase IoTZip's efficiency and discuss the most significant prediction errors that affect its accuracy. We perform experiments in two different ways, (a) under constant network settings and (b) under network throughput that varies over time.

\subsection{Evaluation under fixed network conditions}

Figure 3 presents the comparison of the aforementioned approaches using bandwidth thresholds at 2Mbps, 5Mbps and  10Mbps. The benchmarks used are distributed across four different test sets, Activity Recognition, Air Quality, RToF Measurements and Alexa, each with different compressibility characteristics as described in Section IV. The Alexa datasets are presented in total first, with subsequent graphs separating into the groups A and B previously described. 
The results are presented as relative speedup over the Uncompressed approach, which is always at 1; higher is better. The figure also includes relative standard errors for each of the policies. As expected, the Time Oracle always demonstrates superior performance over the other policies as it represents an ideal scenario in which IoTZip makes decisions correctly across all data. 

As Figure 3 shows, IoTZip performs better relative to the Uncompressed approach when the available bandwidth is low. That is expected since network transfers become gradually more expensive as bandwidth declines. Compression makes better use of limited bandwidth, both in terms of sending less data to begin with, and also in terms of requiring fewer retries. 
 Similarly, IoTZip performs better versus Compressed as the network throughput increases. An always-compress strategy can be inefficient for fast networks, because the additional compression latency can outweigh the benefits of transferring less data. IoTZip shows performance advantages against both the Compressed and Uncompressed approaches.
 
 Apart from the network conditions, the type of data transferred also affects the performance of each policy. Highly compressible data favor the Compressed method, as the use of compression allows to pay overhead to reduce the size of data significantly. However, when data are not compressible, the Uncompressed approach has better performance since introduction of compression adds overhead but yields limited data savings in return. IoTZip benefits in both scenarios as it can make a data driven decision. It achieves a maximum speedup of 3.78x, whereas the average speedup across datasets is at 2.18x.
 
For the highly compressible data in the Activity Recognition and RToF datasets, there are some cases where the Compressed policy edges IoTZip, whereas in the Air Quality dataset that is highly non compressible, Uncompressed is marginally better than IoTZip.
 
 Since the Alexa dataset includes data with a wide spectrum of data size and compressibility, it allows us to better observe the aforementioned tradeoffs and we will study it in more detail in Figures 4(a)-4(f). Here, since the Alexa dataset is divided into two test sets according to compressibility, we can compare IoTZip's advantages in two distinct scenarios. Each of these test sets is comprised by 25 mobile website benchmarks, each corresponding to mobile web data contained in the page load for that particular website. The benchmarks are sorted in descending order of data compressibility. For each of these benchmarks we present the relative speedup of the Compressed, IoTZip and Time Oracle policies normalized over the Uncompressed policy. Additionally, we present the average of these benchmarks for each of the two test sets A and B and across the network conditions 2, 5 and 10 Mbps.
 
 Looking at the averages for each figure we conclude that Test set A (more compressible) provides better performance than test set B, since there are more opportunities to reduce network transfer times when utilizing compression. Test set B contains data with low compressibility, which IoTZip often chooses to leave uncompressed, and often due to the file size and type threshold criteria. This approach proves to be beneficial in the case of low throughput at 2 Mbps and 5 Mbps. When throughput is low, compressing is the common case, as savings from network transfers are more significant. Therefore, IoTZip makes fewer errors on test set B and has less slowdown against the Oracle compared to test set A. However, that does not hold for 10 Mbps, where IoTZip performs marginally better on test set A. 

\begin{table}[t]
\centering 
\caption{Overview of performance and accuracy statistics for IoTZip across different network conditions for the Alexa datasets. Comparison to the {\em Uncompressed}, {\em Compressed} and {\em Time Oracle} approaches presents the percentage of benchmarks (data corresponding to individual mobile web pages) sped up with IoTZip, the respective average speedup and the library's overhead.}
\begin{tabular}{ ||l | c | c | c|| }
 \hline & 
   \multicolumn{3}{|c|}{\textbf{Network conditions}}  \\
   \hline
& \textbf{2 Mbps} & \textbf{5 Mbps} & \textbf{10 Mbps} \\
 \hline
 Percentage win over Compressed &76\% &86\%& 86\% \\
 \hline
 Percentage win over Uncompressed &92\% &88\%& 90\% \\
 \hline
 Speedup vs Uncompressed & 1.78 &1.56 &1.53\\
 \hline
 Speedup vs Compressed &1.12 &1.19&1.24\\
 \hline
 IoTZip overhead &0.26\% & 0.36\% & 0.56\%\\
 \hline
 Slowdown vs Oracle &0.84 &0.81 &0.86\\
 \hline
\end{tabular}
\label{tab:stats1}
\end{table}

Table \ref{tab:stats1} presents statistics about IoTZip's performance across different network settings using the Alexa dataset. First, the table presents the percentage of Alexa website benchmarks for which IoTZip outperforms the Compressed and Uncompressed approaches. The table presents results at a per-benchmark granularity, comparing the aggregate network transfer time of each website's data for every policy. In addition, the table depicts the average speedup of IoTZip across both test sets for the Uncompressed and Compressed approaches and its performance relative to the Time Oracle. It also presents IoTZip's average overhead, which ranges from 0.26\% to 0.56\% of the total transfer time of a request. 
To better understand how IoTZip resolves the tradeoffs we break down the different latencies involved in each request performed by IoTZip in Figure 5. The Total Runtime is composed by the Overhead, Compression Time and Transmission time, in order to process, possibly compress and transfer the data. The latencies are presented normalized to the Total Time along with their respective relative standard errors. We observe how IoTZip is affected by the changes in network conditions. The results at 2 Mbps show that most of the latency is taken up by the Transmission time and as the network throughput increases the Transmission Time decreases and the Compression Time increases relative to the Total Time. This behavior is expected as the compression is constant across network conditions. As the network throughput increases, the compression overhead becomes increasingly significant until a tipping point where compression no longer provides enough data savings and reduction in data transmission time. Data in the Activity Recognition and RToF datasets are highly compressible and the percentage of Total Time used for Compression is high compared to the Alexa dataset where data is less compressible. This happens as IoTZip chooses to compress more frequently as compression is more beneficial. In addition, the data savings are much higher in the compressible datasets which corresponds to shorter overall Transmission Time. For the Air Quality dataset there is no compression as IoTZip's threshold is in force due to the small data size.

IoTZip is designed to inherently provide data savings, as it bases its approach on compression. We measured the data usage for each policy across different network conditions and across all datasets and present them in Figure 6. The data usage for each policy is presented normalized over the data usage of the Uncompressed policy and lower is better. The Uncompressed policy is guaranteed to have the highest data usage as there is no compression involved. Additionally, the Time Oracle, despite providing ideal performance, does not necessarily provide the most data savings, as in some transfers the correct decision is to leave data uncompressed. The data usage demonstrates variation across the different network settings, as IoTZip and the Time Oracle can become more aggressive on using compression if the network throughput is reduced and less aggressive if the throughput is high. The figure additionally shows the data usage based on the different datasets. Having highly compressible data can drastically change the data usage as is demonstrated by the Activity Recognition and RToF Measurements dataset. After transferring the former, IoTZip reduces data Usage to 24.2\% of the original and in the case of the latter data usage goes down to 18.5\%. In the case of the Air Quality dataset, the threshold disallows compression and therefore the data usage remains unchanged. For the Alexa dataset, data usage is at roughly 60\% of the original at 2, 5, and 10Mbps, while providing applications with performance speedup. The majority of data savings originates from large transfers of highly compressible data. For all datasets, IoTZip performs very well in terms of data savings and is operating close to the optimal, as demonstrated by the Compressed policy columns.

IoTZip's performance compared to other policies correlates with its prediction accuracy, as depicted in Table IV. These results demonstrate the accuracy of IoTZip across different network conditions for all datasets. For this comparison, the Time Oracle is used as ground truth. IoTZip is subject to two different kinds of errors: when it decides to compress when it should not (false positives) and when it fails to identify that compression is beneficial (false negatives). Most incorrect predictions are false negatives. The false negative rate in the Alexa dataset decreases as the throughput increases, because compression is no longer beneficial for some of the data. The same trend does not hold in the Air Quality dataset, as IoTZip's threshold does not allow compression due to small data size. Activity Recognition and RToF Measurements datasets have better success rate since large file size and highly compressible data make the compression decisions easier.  
\begin{table}[t]
\label{tab:stats2}
\centering 
\caption{Characterization of the most significant prediction errors across all datasets for IoTZip using the Time Oracle as ground truth.}
\begin{tabular}{ ||l | c | c | c | c|| }
 \hline
  \textbf{Dataset}&\textbf{Error Type} & \textbf{2 Mbps} & \textbf{5 Mbps} & \textbf{10 Mbps} \\
 \hline
Activ. R. & Success Rate (\%)&100 &100 &98.67\\
 \hline
 & False positives (\%)&0 &0&1.33\\
 \hline
 & False negatives (\%)&0 &0 &0\\
 \hline
 Air Q. & Success Rate (\%)&65.00 &80.88 &74.84\\
 \hline
 & False positives (\%)&0 &0  &0\\
 \hline
 & False negatives (\%)&35.00 &19.12 &25.16\\
 \hline
 RToF M. & Success Rate (\%)&100 &87.33 &92.66\\
 \hline
 & False positives (\%)&0 &12.67  &7.33\\
 \hline
 & False negatives (\%)&0 &0 &0\\
 \hline
 Alexa &Success Rate (\%)&74.66&74.36 &77.07\\
 \hline
 & False positives (\%)&3.41 &4.90 &4.34\\
 \hline
 & False negatives (\%)&21.93&20.74&18.59\\
 \hline
\end{tabular}
\end{table}

Our evaluation of IoTZip under fixed network conditions indicates that it performs consistently better than the Uncompressed and Compressed policies and approaches the ideal performance (Time Oracle) in many cases. IoTZip demonstrates an average speedup of 2.18x over the Uncompressed policy across all datasets, with a maximum of 3.78x. It does so while maintaining high accuracy throughout all datasets and while demonstrating significantly reduced data usage down to 18.5\% of the original data size. 

\begin{figure}
 \centering
  \includegraphics[width=1\linewidth]{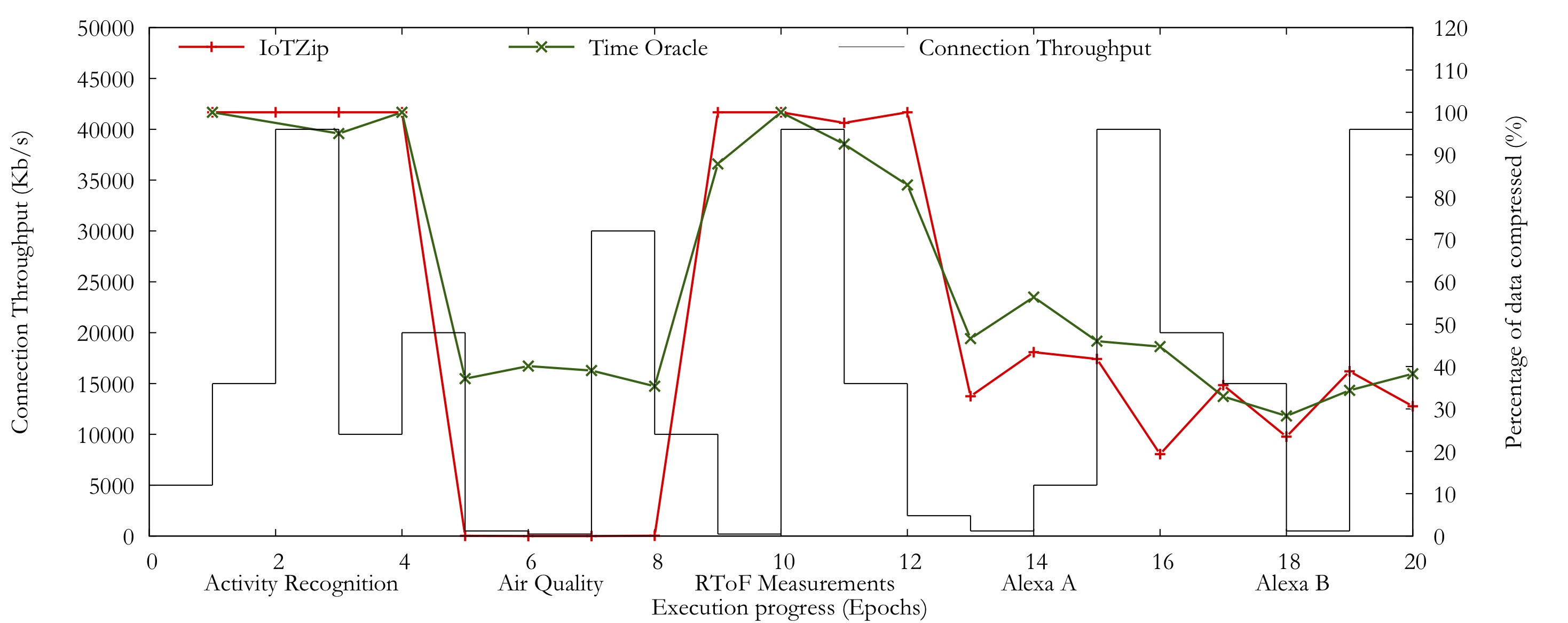}
  \label{fig:results}
  \vspace{-10pt}
   \caption{Presentation of the percentages of compressed data over time for IoTZip and Time Oracle. Results are presented across each epoch along the network throughput level at each epoch. The changes in percentage of compressed data and IoTZip's ability to "follow" the Time Oracle indicate IoTZip's adaptivity to changes in network conditions.}
\end{figure}

\begin{figure}
 \centering
  \includegraphics[width=1\linewidth]{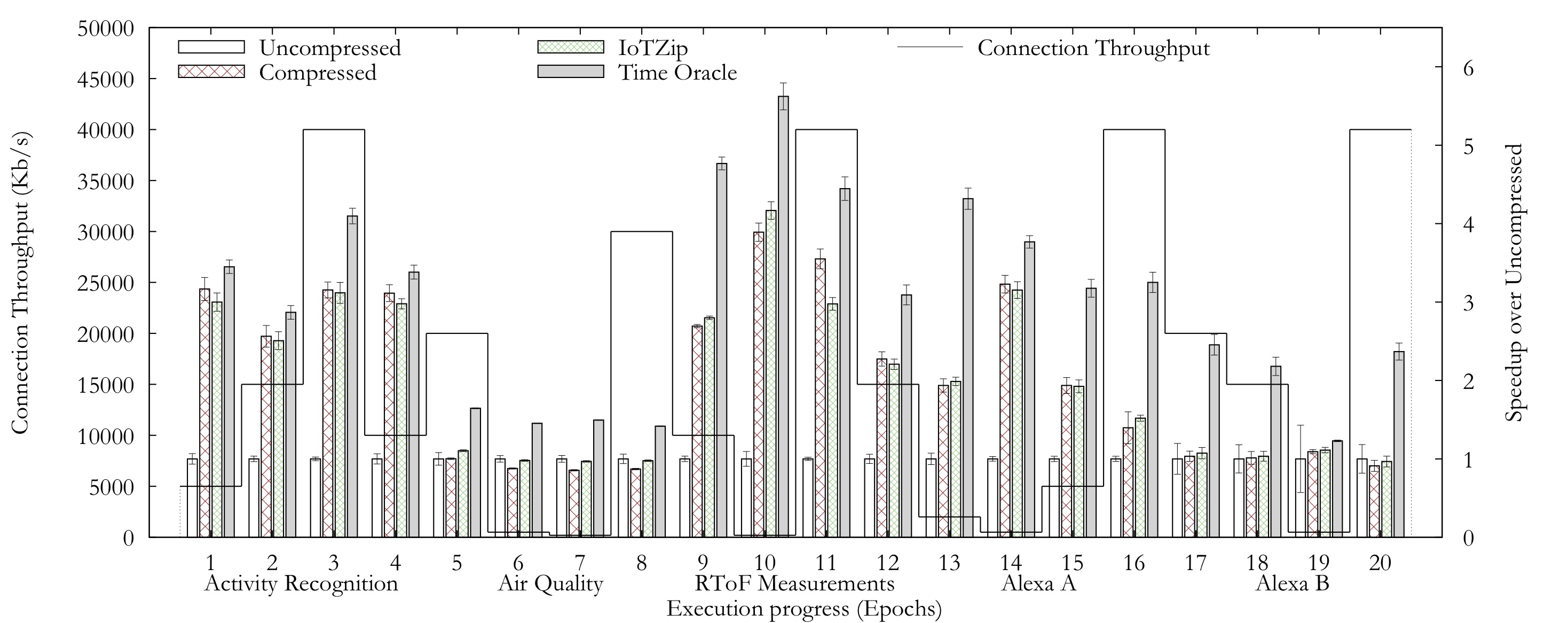}
  \label{fig:results}
  \vspace{-10pt}
   \caption{Speedup of compression policies across datasets and network settings. Results are presented across each epoch along the network throughput level at each epoch. Results are presented relative to the Uncompressed policy along with their respective standard errors.}
\end{figure}

\subsection{Evaluation under changing network conditions}

Having seen IoTZip's speedup advantages for constant network conditions, we next extend our evaluation to demonstrate that IoTZip can offer performance improvements while adapting to changes in network conditions. 
Figure 7 demonstrates the percentage of compressed data for IoTZip and the Time Oracle across the timeline. The figure also depicts the changes in network throughput across the timeline. For our experiments we use the IoT datasets and Alexa test sets in sequence to build the timeline. Each dataset is split in 4 partitions and results are presented per partition. The transfer of each partition corresponds to an epoch. During the timeline, changes in network throughput happen across epochs, but network settings remain constant within an epoch. The Time Oracle line presents the percentage of data compressed under the assumption that perfect compression decisions are made. 

In the figure we can observe the fluctuations in the rate IoTZip performs correct compression decision predictions over time. We observe that the percentage of compressed data changes over the course of the experiment and is affected by the data and the network throughput. The IoTZip line demonstrates the same behavior as the Time Oracle, although with some false positives and false negatives. As the network throughput decreases, the percentage of compressed files increases and the trend occurs reversed during an increase in the available network throughput. Although each change in network throughput triggers a change in the percentage of data compressed, this percentage is affected by the size of data in the dataset and their compressibility. When comparing epochs around the RToF Measurements and the Alexa B datasets, we can observe that in the first case changes in the network throughput cause the percentage to fluctuate around 100\%, whereas in the latter case the percentage fluctuates between 20-40\%. One exception is the Air Quality dataset. Due to their very small size, the data don't pass the IoTZip's threshold and are immediately disqualified for compression. Therefore, there is a large gap between the IoTZip and Time Oracle lines.

IoTZip's throughput prediction follows the throughput movement the device observes and has a trend similar to the Time Oracle, although there is a gap between the two lines. One reason for that gap is that there is a disparity between IoTZip's throughput prediction and the real bandwidth. First,  IoTZip's throughput prediction module uses a moving average that adjusts the prediction progressively and smooths rapid changes in measured throughput. The reason for this mechanism is to avoid the throughput estimate to fluctuate significantly during rapid changes in throughput.
The average prediction accuracy is 74.04\% across the timeline. Table V presents an error characterization for IoTZip across the experiment timeline. The errors are evaluated against the Time Oracle. We can observe that the success rate of highly compressible datasets (Activity Recognition and RToF Measurements) is high and in this case most errors fall under false positives and IoTZip compresses when it should not. Epochs 5-9 that belong to the Air Quality dataset only have false negatives and low success rate, as IoTZip's threshold limits compression on these data. When comparing the success rate between Alexa A and Alexa B datasets, we observe that similarly Alexa A has a better success rate due to its higher compressibility. A technique capable of adapting the threshold based on the data and network conditions could be investigated in future work.

\begin{table*}[t]
\label{tab:stats2}
\centering 
\caption{Characterization of the most significant prediction errors for IoTZip using the Time Oracle as ground truth. The prediction errors are presented per epoch and network throughput varies across epochs.}
\setlength\tabcolsep{3.8pt}
\begin{tabular}{ |*{21}{c|}}
 \hline
  \textbf{Epoch} & 1 & 2 & 3 & 4 & 5 & 6 & 7 & 8 & 9 & 10 & 11 & 12 & 13 & 14 & 15 & 16 & 17 & 18 & 19 & 20 \\
 \hline
 Success Rate (\%)   &100 &100 &95.0 & 100 & 62.9 &59.9 & 60.9 & 64.7 & 87.8 & 100  & 90.0  & 82.9 & 75.6 &74.2  & 69.4  & 60.6 & 62.1  & 67.7 & 67.3 & 59.8 \\
 \hline
 False positives (\%) &0 &0 &5.0 & 0 & 0  & 0& 0 & 0  & 12.2 & 0 & 7.5 & 17.1 & 5.4 & 6.4 & 13.2 & 7.0 & 20.3 & 13.7  & 18.6  & 16.3 \\
 \hline
 False negatives (\%) & 0& 0&0 &0  & 37.1 & 40.1 & 39.1 & 35.3 & 0 & 0 & 2.5 & 0 & 19.0 & 19.4  & 17.4  & 32.4 & 17.6  & 18.6 & 14.1  & 23.9 \\
 \hline
\end{tabular}
\end{table*}

Figure 8 presents the relative speedup of the Compressed, IoTZip and Time Oracle policies over the Uncompressed across the epochs of the timeline. The network throughput level for each epoch is also available for each figure. The effect of the changes in network conditions are apparent on the results. Whenever the network throughput increases, we observe that the IoTZip performs better compared to the Compressed version and worse compared to the Uncompressed version. Similarly, the reversed behavior occurs when the network throughput is low. Based on the compressibility of the datasets, IoTZip provided better speedup when data are highly compressible. However, in some cases the Compressed policy edges IoTZip. When the data is not compressible as in the Air Quality dataset, IoTZip always performs better than the Compressed policy but can marginally perform worse than the Uncompressed policy. As previously mentioned, change of network conditions affects the prediction accuracy since IoTZip uses an approach that progressively propagates the changes in throughput to the model. Therefore, changes in network conditions affect, even temporarily, the accuracy of compression decisions. As evidenced by comparing tables IV and V, IoTZip's prediction errors are on average better under fixed network settings when compared against measurements performed under dynamically changing network conditions.

We evaluated IoTZip in dynamically changing network settings across all datasets where our library proved to be superior than the uniform policies, Uncompressed and Compressed. Our results demonstrate a maximum speedup of 4.17$\times$ and an average speedup of 2.03$\times$  over the Uncompressed policy. IoTZip is adaptive to changes in network conditions and follows the Time Oracle’s  behavior while in many cases its performance is close to ideal. IoTZip showcases an average prediction accuracy of 74.04\% despite the varying network conditions.

\section{Conclusion}
This paper presented IoTZip, a library for optimizing IoT and mobile web traffic which implements selective compression---using it only when it is likely to benefit performance. To support this, IoTZip uses compression latency and network throughput estimates to reason about the compression decision of each web transfer. 

Based on our analysis of IoT data, specific IoT systems operate with a single type of data and can generate data that exhibit little variation in data size. Although these data characteristics could be used as indications to reduce the need for compression selectivity, throughput and network quality variations will still be present in IoT environments, making compression selectivity necessary.

Throughout the evaluation of IoTZip, it is clear that the performance and data usage of IoT communication heavily relies on data characteristics. We envision that our work can also be utilized as a tool to characterize IoT applications. IoTZip can identify properties of application data (data size distribution, compressibility) as well as provide insight on how to handle communication efficiently based on a Time Oracle that documents correct compression decisions for the data and the network conditions in question. 

IoTZip performs consistently better than uniform policies requiring either all-compressed or all-uncompressed data and also approaches the Time Oracle policy in many cases . Its average prediction accuracy is above 70\% and its resulting runtime latency outperforms these na{\"i}ve policies, delivering a speedup of up to  3.78x. The average speedup of IoTZip is 2.18x and 2.03x across datasets under fixed and dynamic network conditions respectively.  Furthermore, the library provides significant data savings across different network conditions and over different data. 

Overall, IoTZip represents an important building block towards broader implementation of traffic-reduction techniques that can improve latency, save energy, and reduce the bandwidth requirements for mobile applications and devices.

%
\bibliographystyle{abbrv}
\bibliography{sigproc}  

\end{document}